\documentclass[aps,prl,preprint,showpacs,superscriptaddress,preprintnumbers,amsmath,amssymb]{revtex4}

\usepackage{graphicx}
\usepackage{dcolumn}
\usepackage{bm}
\newcommand{\ve}{\boldsymbol}

\begin{document}

\preprint{APS/123-QED}

\title{Active control of light trapping by means of local magnetic coupling}%

\author{M. Burresi}
\email[]{burresi@amolf.nl,burresi@lens.unifi.it}
\affiliation{Center for Nanophotonics, FOM Institute for Atomic and
Molecular Physics (AMOLF), Science Park 113, 1098 XG Amsterdam, The
Netherlands}

\author{T. Kampfrath}

\affiliation{Center for Nanophotonics, FOM Institute for Atomic
and Molecular Physics (AMOLF), Kruislaan 407, 1098 SJ Amsterdam,
The Netherlands}

\author{D. van Oosten}

\affiliation{Center for Nanophotonics, FOM Institute for Atomic
and Molecular Physics (AMOLF), Science Park 113, 1098 XG Amsterdam,
The Netherlands}

\author{J. C. Prangsma}

\affiliation{Center for Nanophotonics, FOM Institute for Atomic
and Molecular Physics (AMOLF), Science Park 113, 1098 XG Amsterdam,
The Netherlands}

\author{B. S. Song}

\affiliation{Department of Electronic Science and Engineering, Kyoto University, Kyotodaigaku-Katsura, Nishikyo-ku, Kyoto 615-8510, Japan.}
\affiliation{School of Information and Communication, Sungkyunkwan University, Janan-Gu, Suwon 440-746, Korea.}

\author{S. Noda}

\affiliation{Department of Electronic Science and Engineering, Kyoto University, Kyotodaigaku-Katsura, Nishikyo-ku, Kyoto 615-8510, Japan.}

\author{L. Kuipers}
\affiliation{Center for Nanophotonics, FOM Institute for Atomic
and Molecular Physics (AMOLF), Science Park 113, 1098 XG Amsterdam,
The Netherlands}

\date{\today}

\newcommand{\PCW}{photonic crystal waveguide}
\newcommand{\POS}{polarization singularity}
\newcommand{\POSs}{polarization singularities}
\newcommand{\PNF}{phase-sensitive near-field microscope}

\begin{abstract}
The ability to actively tune the properties of a nanocavity is
crucial for future applications in photonics and quantum
information. Two important man-made classes of materials have
emerged to mold the flow of electromagnetic waves. Firstly, photonic
crystals are dielectric nanostructures that can be used to confine
and slow down light and control its emission. They act primarily on
the electric component of the light field. More recently, a novel
class of metallo-dielectric nanostructures has emerged. These
so-called metamaterials enable fascinating phenomena, such as
negative refraction, super-focusing and cloaking. This second class
of materials realizes light control through effective interactions
with both electric and magnetic component. In this work, we combine
both concepts to gain an active and reversible control of light
trapping on subwavelength length scales. By actuating a nanoscale
magnetic coil close to a photonic crystal nanocavity, we interact
with the rapidly varying magnetic field and accomplish an
unprecedented control of the optical properties of the cavity. We
achieve a reversible enhancement of the lifetime of photons in the
cavity. By successfully combining photonic crystal and metamaterials
concepts, our results open the way for new light control strategies
based on interactions which include the magnetic component of light.
\end{abstract}


\maketitle

Photonic crystals are materials which provide a high level of
control on the light-matter interaction, based on the engineered
periodic modulation of the electric permittivity
\cite{yablonovitch_inhibited_1987}. Nanoresonators in such
photonic crystal architectures can store light in volumes
comparable to the wavelength cubed for times longer than a million
oscillation periods of the light \cite{takahashi_high-q_2007}.
Such high-$Q$ photonic nano-cavities are promising structures to
achieve strong coupling between light and quantum dots
\cite{yoshie_vacuum_2004,hennessy_quantum_2007}. The ability to
actively tune the properties of a nanocavity is crucial for future
applications in photonics and quantum information
\cite{yoshie_vacuum_2004,hennessy_quantum_2007}. Active tuning is
achievable all-optically \cite{almeida_all-optical_2004},
electrically \cite{xu_micrometre-scale_2005} or through the
actuation of nano-objects in the evanescent electromagnetic field
of the cavity
\cite{koenderink_controllingresonance_2005,hopman_nano-mechanical_2006,lalouat_near-field_2007,mujumdar_near-field_2007,intonti_spectral_2008}.
The latter strategy, which could lead to breakthroughs in the
emerging field of optical nanoelectromechanical systems (NEMS)
\cite{li_harnessing_2008}, relies typically on the interaction
with the electric field in the cavity. This invariably leads to
only red shifts of the resonance frequency and usually to a
reduction of the photon lifetime of the cavity
\cite{koenderink_controllingresonance_2005,hopman_nano-mechanical_2006,lalouat_near-field_2007,mujumdar_near-field_2007,intonti_spectral_2008}.
In principle, an interaction with the magnetic field would also
allow tuning of the cavity \cite{lalouat_near-field_2007}.
Unfortunately, natural materials have a negligible magnetic
permeability at optical frequencies. We can overcome this
limitation by borrowing concepts from metamaterials. These
engineered materials work by geometrically inducing a magnetic
response \cite{soukoulis_physics:_2007}. By using this idea, we
have achieved active and reversible tuning of a photonic crystal
nanocavity by interacting with the magnetic field of the trapped
light. We use a cylindrically symmetric, metal-coated probe as a
'nanocoil'. By positioning the probe close to the cavity, the
z-component of the magnetic field induces a counteracting magnetic
response in the nanocoil through Lenz' law. As a result, we are
able to induce a novel blue shift of the resonance frequency. More
importantly, we are able to achieve an increase of the quality
factor $Q$. In other words, we are able to increase the photon
lifetime in the cavity.
\begin{figure} [p]
\begin{center}
\includegraphics[width=12cm]{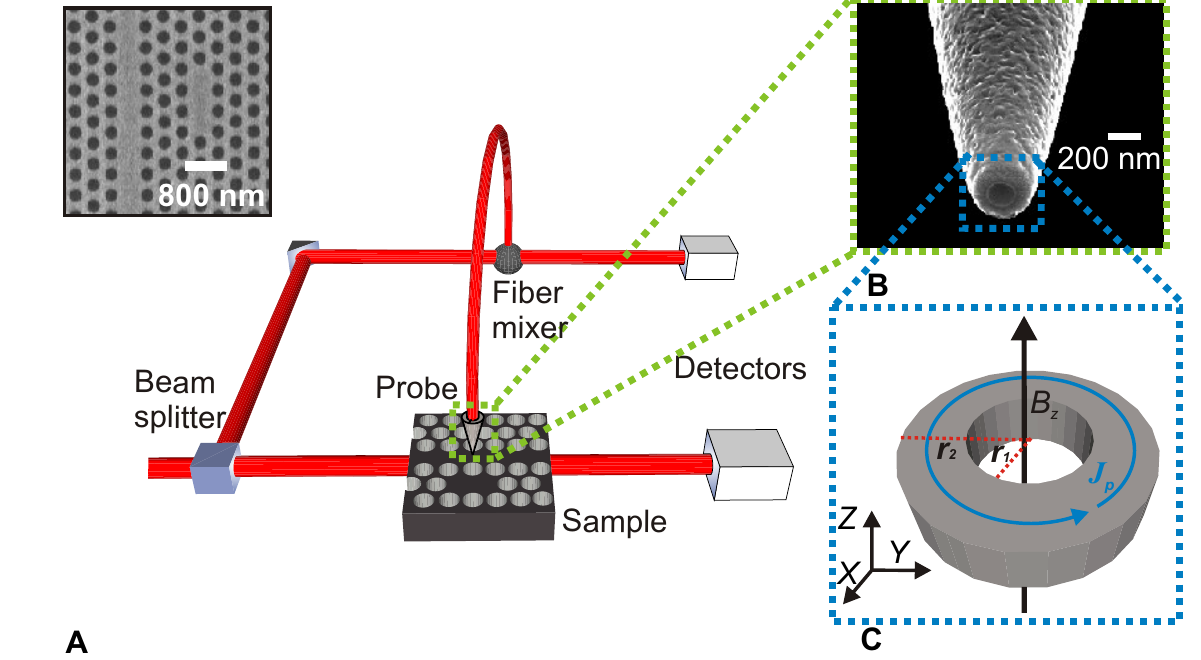}
\caption{{\rm \textbf{A}},The near-field probe is scanned above
the sample and collects the evanescent field of the light in the
structure. The collected light is mixed with a reference beam and
subsequently detected with a heterodyne scheme. The light power
transmitted by the structure is also detected. A scanning electron
micrograph of the sample investigated is shown in the inset. The
photonic crystal nanocavity is visible below the photonic crystal
waveguide. {\rm \textbf{B}}, A scanning electron micrograph of the
cylindrical symmetric aluminium-coated near-field probe. {\rm
\textbf{C}}, Schematic representation of the ring that models the
end of the near-field probe. The magnetic field $B_z$, that is
orthogonal to the ring, induces a current density $J_p$ in the ring.
$r_1$ and $r_2$ are the outer and the inner radius,
respectively.}\label{fig4.1}
\end{center}
\end{figure}

We investigate a photonic crystal nanocavity which is side-coupled
to a photonic crystal waveguide (see inset Fig. \ref{fig4.1}A) \cite{akahane_high-q_2003}. The
\begin{figure} [t]
\begin{center}
\includegraphics[width=12cm]{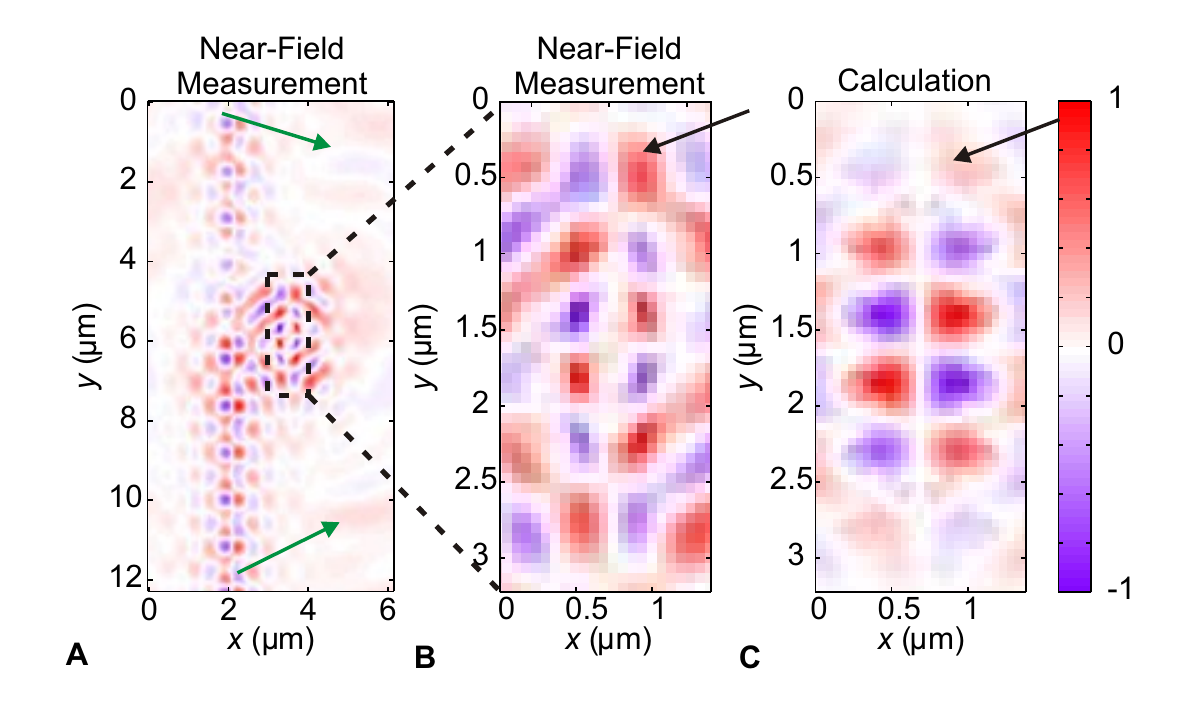}
\caption{{\rm \textbf{A}}, Distribution of Re$(E_y)$ detected in
the near field of the sample at resonance. The electromagnetic field
depicted here is propagating through the accessing waveguide and
coupled to the cavity. The color scale is varying between the
maximum (red-positive) and the minimum (violet-negative) of the
interference signal. {\rm \textbf{B}}, Image of the distribution of
the detected Re$(E_y)$ in the near field above the area indicated in
the dashed box of Fig. \ref{fig4.2}A. The green arrows indicate the
detected in-plane radiation lost by the cavity. {\rm \textbf{C}},
Image of the calculated distribution of the longitudinal component
obtained by FDTD calculations at resonance.} \label{fig4.2}
\end{center}
\end{figure}
nanocavity exhibits a resonance at a vacuum wavelength of
$\lambda_\mathrm{o} = 1534.6$ nm with a quality factor $Q_\mathrm{o}
= 6500$. In order to excite the cavity, light from a tunable diode
laser is coupled to the access waveguide. The electric field
distribution inside and around the cavity is detected with a
phase-sensitive near-field microscope
\cite{balistreri_tracking_2001} (Fig. \ref{fig4.1}A). By raster
scanning a tapered aluminum-coated single-mode fibre (Fig.
\ref{fig4.1}B) above the sample at a constant height of 20 nm, we
collect a minute fraction of the light and detect it with a
heterodyne scheme. The near-field probe, which has an aperture of
200 nm and an aluminum coating of 100 nm, has a cylindrical symmetry
\cite{veerman_high_1998}. The high symmetry of the probe allows us
to detect the in-plane electric field distribution of the sample
\cite{burresi_observation_2009}. Figure
\ref{fig4.2}A displays the distribution of Re$(E_y)$ detected with a
typical near-field measurement at vacuum wavelength
$\lambda_\mathrm{o}$. The image shows how the electromagnetic wave
is guided by the waveguide and is coupled to the cavity, which is
indicated by the dashed box in Fig. \ref{fig4.2}A. Excellent
agreement is found between the measured (\ref{fig4.2}B) and the
calculated (\ref{fig4.2}C) field distribution above the cavity,
obtained by Finite Difference Time Domain (FDTD) method. Small
deviations between theory and experiment are visible. The field
outside the cavity appears stronger in the measurement than in the
calculation, as indicated by the arrows. We assign this effect to
the influence of the probe on the optical properties of the cavity,
as it will been described later in this chapter.

While performing the near-field measurement, we simultaneously
determine the transmission of the system by measuring the amount
of light arriving at the output of the access waveguide. In order
to investigate the influence of the probe on the transmittance of
the system, we determine the normalized transmission $F(x, y,
\lambda) = T_\mathrm{n}(x, y, \lambda)/T_\mathrm{o}(\lambda)$,
where $x$ and $y$ represent the in-plane position of the probe,
$T_\mathrm{n}(x, y, \lambda)$ is the transmission spectrum as a
function of probe position above the cavity, and
$T_\mathrm{o}(\lambda)$ is the unperturbed transmission spectrum
obtained in absence of the probe.
\begin{figure} [t]
\begin{center}
\includegraphics[width=12cm]{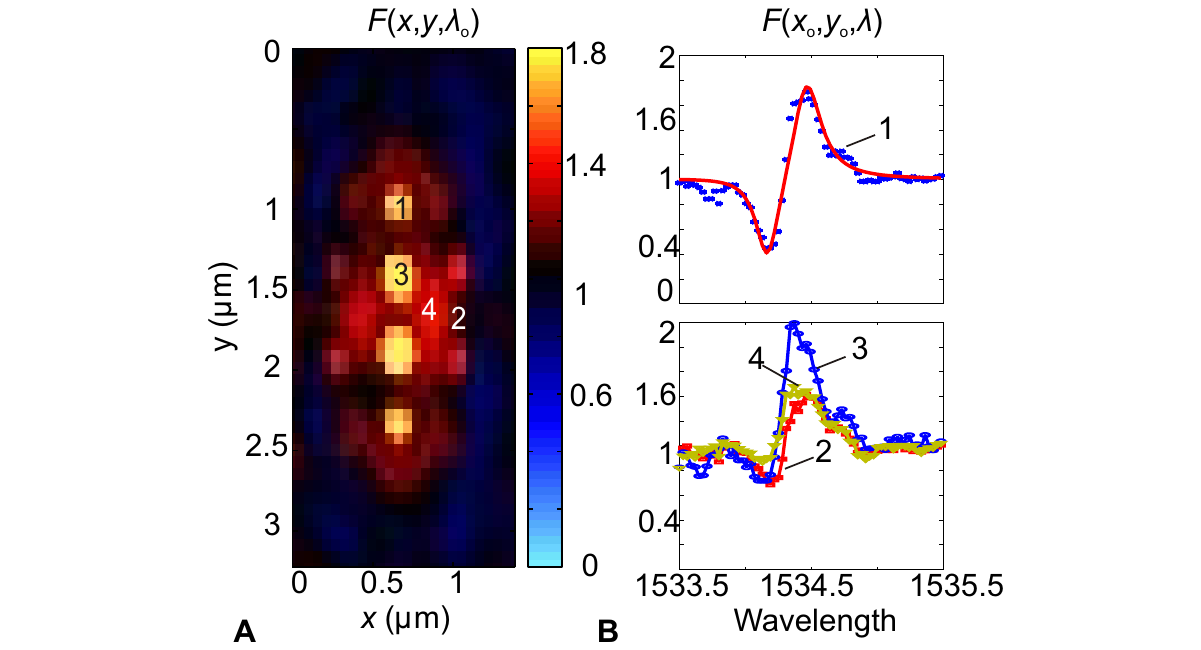}
\caption{{\rm \textbf{A}}, Image of the normalized transmission
at resonance. In black the areas where the transmittance equals
the unperturbed transmittance. For different positions of the
probe the transmittance either increases or decreases. {\rm
\textbf{B}}, In the upper image the normalized transmission $F$
for position 1 in Fig. \ref{fig4.3}A is shown. For wavelengths far
away from the resonance the ratio is 1, indicating no influence of
the probe. However, for wavelengths close to the resonance this
ratio varies drastically. The fit is shown as a red line. The
lower image shows $F$ obtained for different positions indicated
in Fig. \ref{fig4.3}A.} \label{fig4.3}
\end{center}
\end{figure}
Figure \ref{fig4.3}A shows a typical image of this normalized
transmission acquired on resonance ($\lambda =
\lambda_\mathrm{o}$). Different probe positions may lead to either
an increase (yellow areas in the image, $F > 1$) or a decrease
(blue areas, $F < 1$) of the waveguide transmittance. This
observation clearly indicates an interaction between probe and
cavity. In addition, we analyze the normalized transmission as a
function of wavelength for a fixed position of the probe. The
upper image of Fig. \ref{fig4.3}B shows a typical graph of $F$ for
position 1 in Fig. \ref{fig4.3}A, whereas the lower image shows
$F$ for 3 other positions indicated in Fig. \ref{fig4.3}A. All
spectra exhibit identical qualitative behavior. For wavelengths
far away from the resonance, the influence of the probe is
negligible, i.e. $F = 1$. For wavelengths close to
$\lambda_\mathrm{o}$, the probe-cavity interaction becomes evident
since F undergoes a pronounced variation. The change in
transmission is caused by a shift of the resonance
($\lambda_\mathrm{o} \rightarrow \lambda_\mathrm{n}(x, y)$) due to
the probe-cavity coupling
\cite{koenderink_controllingresonance_2005,lalouat_near-field_2007},
as a consequence of which light with wavelength
$\lambda_\mathrm{o}$ no longer couples to the resonator. As a
result, the light does not experience the small loss associated
with being trapped in the cavity and the transmission increases.
Conversely, light with a wavelength close to the new resonance
$\lambda_\mathrm{n}$ is now loaded in the resonator, leading to a
reduction of the transmittance for that wavelength. It is clear
that a distinct blue-shift of the cavity resonance occurs when the
near-field probe couples to the cavity.

The induced resonance shift is the cause of the above-mentioned
disagreement between Fig. \ref{fig4.2}B and C. When the probe is
above the cavity, light with wavelength $\lambda_o$ is not loaded
and the electric field in the resonator is smaller than in the
unperturbed system. Conversely, when the probe is at the position
indicated by the arrow in Fig. \ref{fig4.2}B, light with
wavelength $\lambda_o$ can couple to the nanocavity and the
electric field at that location, as well as the signal
detected by the probe, increases. As a result, the ratio between
the electric field amplitude inside and outside the cavity for the
measurement (Fig. \ref{fig4.2}B) differs from the same ratio for
the calculation (Fig. \ref{fig4.2}C).

This unprecedented blue-shift can be intuitively understood in the
following way. Due to the small extension in air of the evanescent
fields above the cavity \cite{engelen_subwavelength_2009}, the end
of our near-field probe can be modeled as a metallic ring (Fig.
\ref{fig4.1}C) that acts like a nano-coil in the electromagnetic
field above the cavity. Faraday's law
tells us that the magnetic field induces a circular current
density $\ve{J}_\mathrm{p}$ in the ring (Fig. \ref{fig4.1}C). This
current, in turn, generates a magnetic field that, according to
Lenz' law, suppresses the driving field inside the ring
\cite{Landau_book_1984}. The probe, thus, generates a volume where
the total magnetic field is reduced. As a result, the effective
volume occupied by the light stored in the cavity, the so-called
cavity mode volume, is reduced, leading to a resonance shift
towards shorter wavelengths.

\begin{figure} [t]
\begin{center}
\includegraphics[width=12cm]{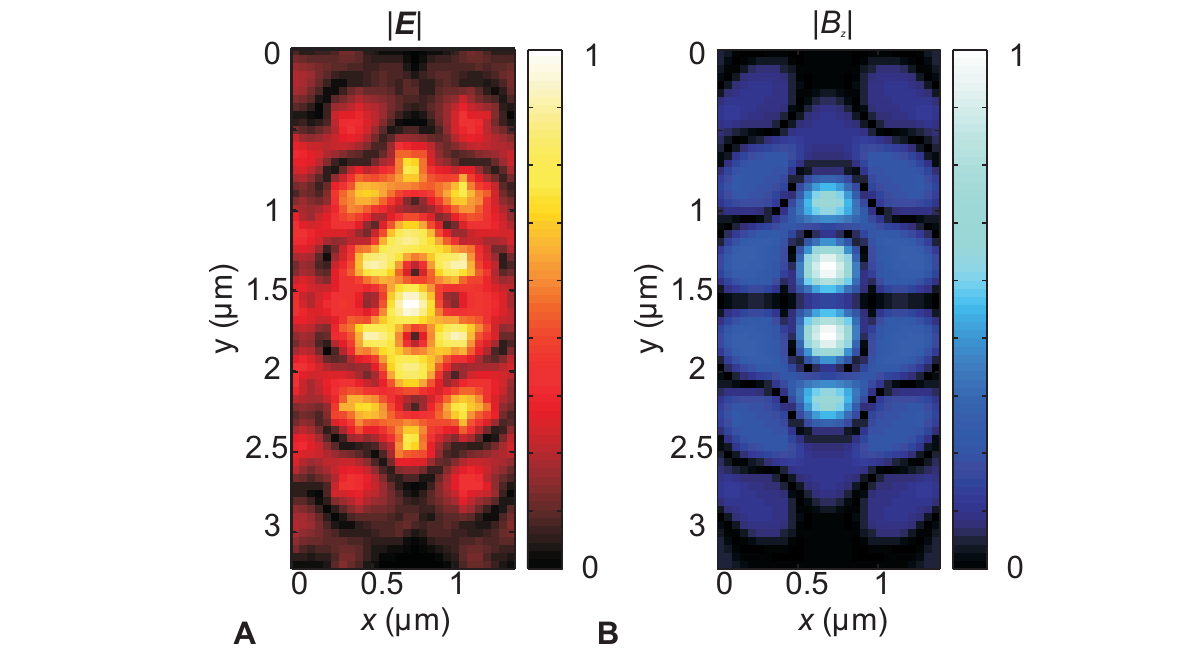}
\caption{{\rm \textbf{A} and \textbf{B}}, Distribution of the
magnitude of the electric field $\ve{E}$ and of the amplitude of
the vertical component of the magnetic field $B_z$ normalized to
their maximum, respectively. The area shown is the same as Fig.
\ref{fig4.3}A. The two pictures show a different symmetry in the
pattern of the field distributions.} \label{fig4.4}
\end{center}
\end{figure}

As a consequence, we expect that the probe-cavity coupling, and thus
the variation of the transmittance, is most pronounced when the
probe overlaps with the maximum in the amplitude of the out-of-plane
component of the magnetic field. We experimentally verify our
expectations by comparing the normalized transmission map, shown in
Fig. \ref{fig4.3}A, with the amplitude distribution of $B_z$,
obtained by FDTD calculation and shown in Fig. \ref{fig4.4}B. The
symmetry and the maxima of the transmission map do coincide with the
amplitude of $B_z$ and not with the magnitude of the electric field
$\ve{E}$ (Fig. \ref{fig4.4}A). This indication proves that the
probe-cavity interaction is dominated by the magnetic coupling.

For a more formal description of the probe-cavity interaction one
has to consider that the relative resonance shift is proportional to
the relative energy shift of the system
\cite{koenderink_controllingresonance_2005,lalouat_near-field_2007}.
Here, we have to take into account the induced magnetic dipole
moment $\ve{m}$ of the probe interacting with the magnetic field in
addition to the coupling between the induced electric dipole moment
$\ve{p}$ of the probe and the electric field. Therefore, the
resonance shift can be written as:
\begin{equation}\label{Eshift}
\frac{\mathrm{\Delta}
\omega_\mathrm{o}}{\omega_\mathrm{o}}=-\frac{\ve{E}_\mathrm{o}^*\cdot\ve{p}+\ve{B}_\mathrm{o}^*\cdot\ve{m}}{2U_E},
\end{equation}
where $\omega_0$ is the resonant angular frequency of the system,
$2U_E$ is the total energy stored in the cavity. The
dipole moments, in turn, are proportional to the unperturbed
$\ve{E}_\mathrm{o}$ and $\ve{B}_\mathrm{o}$ and can be expressed
as $p_i=\alpha^{ee}_{ii}E_i$ and $m_i=\alpha^{mm}_{ii}B_i$, where
the label $i$ indicates the spatial coordinates $x$, $y$ and $z$.
The proportionality constants $\alpha^{ee}_{ii}$ and
$\alpha^{mm}_{ii}$ are the electric and magnetic polarizabilities
of the probe, respectively. From eq.
\ref{Eshift} it is clear that when the perturbative object only
exhibits an electric response, the transmittance variation were
largest when the probe overlaps with the amplitude maxima of the
electric field
\cite{koenderink_controllingresonance_2005,hopman_nano-mechanical_2006,lalouat_near-field_2007,mujumdar_near-field_2007,intonti_spectral_2008}.
In order to calculate the electric polarizabilities, we can
approximate the ring as a metallic oblate spheroid, following the
methodology often employed in split-ring resonators
\cite{marques_role_2002}. The electric polarizabilities turn out
to be positive. The magnetic
polarizability can be calculated by applying Faraday's law to a
single metallic loop. This leads to a negative polarizability
$\alpha^{mm}_{zz}=-{A^2/(L+iR/\omega)}$, where $L$ is the
self-inductance of the ring, $R$ is the complex Ohmic resistance
and $r_2$ is the outer radius. Besides
exhibiting a positive electric polarizability, our near-field
probe also has a negative magnetic polarizability. Thus, the
electric coupling $\ve{E}_\mathrm{o}\cdot \ve{p} > 0$ induces a
red-shift
\cite{koenderink_controllingresonance_2005,hopman_nano-mechanical_2006,lalouat_near-field_2007,mujumdar_near-field_2007,intonti_spectral_2008},
whereas the magnetic coupling $\ve{B}_\mathrm{o}\cdot \ve{m} < 0$
leads to a blue-shift. Thus, when the magnetic coupling dominates
the resonance is primarily blue-shifted.

\begin{figure} [t]
\begin{center}
\includegraphics[width=12cm]{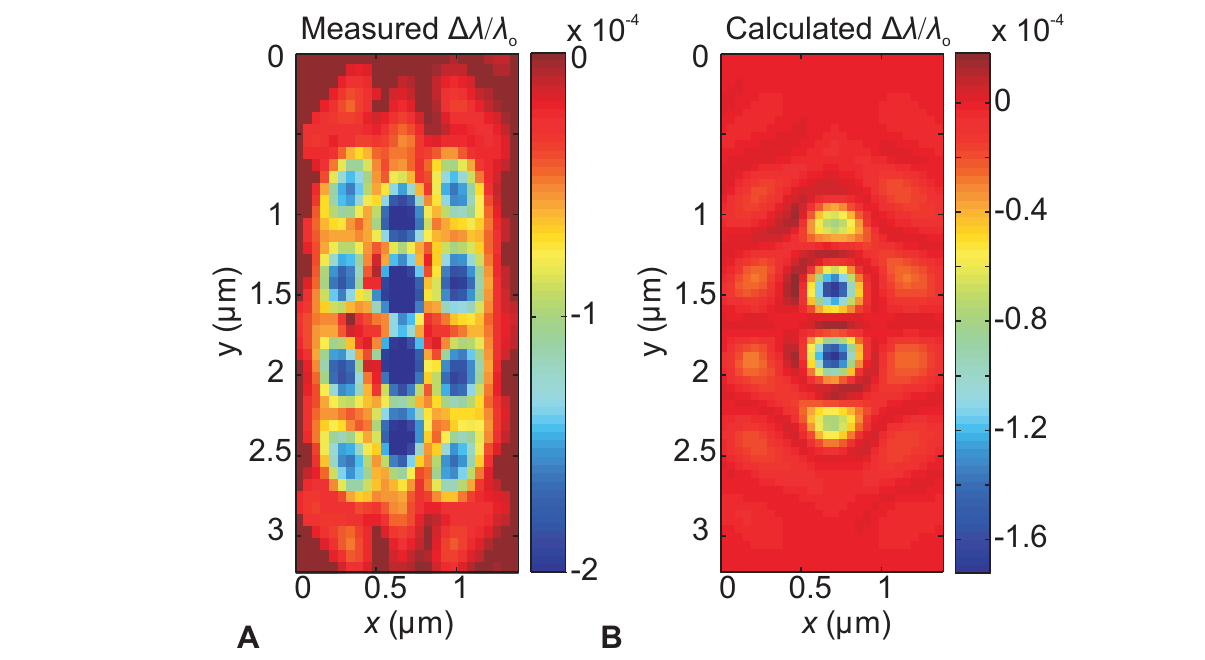}
\caption{{\rm \textbf{A} and \textbf{B}}, Images of the measured
and calculated shift of the resonance
$\mathrm{\Delta}\lambda_\mathrm{o}/\lambda_\mathrm{o}$ at every
probe position. Figure \ref{fig4.5}A shows an evident blue-shift of
the resonance when the probe is above maxima of the $|B_z|$. We find
an excellent quantitative agreement with the calculated resonance
shift in Fig. \ref{fig4.5}B.} \label{fig4.5}
\end{center}
\end{figure}

In order to compare the theoretical prediction of eq. \ref{Eshift}
to our experimental data, we extract the relative shift
$\mathrm{\Delta}\lambda/\lambda_\mathrm{o}$ from our measurements.
For this purpose, we fit a transmission function based on
coupled-mode theory \cite{manolatou_coupling_1999} to the normalized
transmission. Figure \ref{fig4.3}B shows a
typical fit as a red line. From the fit we obtained the relative
resonance shift $\mathrm{\Delta}\lambda/\lambda_\mathrm{o}$ and the
relative change in the quality factor
$\mathrm{\Delta}Q/Q_\mathrm{o}$ for all probe positions. In Fig.
\ref{fig4.5}A we show the measured
$\mathrm{\Delta}\lambda/\lambda_\mathrm{o}$ as a function of the
probe position. By comparing with Fig. \ref{fig4.3}A, it is evident
that the largest relative blue-shifts, of the order of $10^{-4}$,
occur for positions of the probe where the amplitude of $B_z$ is
maximum. We calculate the resonance shift by using eq. \ref{Eshift}
and the field distributions inside the cavity, which were obtained
by FDTD calculations. In order to take into account the finite size
of the probe, we use the average electric and magnetic fields over
the area of the ring by making a convolution of the probe apex shape
with the calculated field distributions. The theoretically obtained
$\mathrm{\Delta}\lambda/\lambda_\mathrm{o}$ is shown in Fig.
\ref{fig4.5}B. We find an excellent qualitative agreement with the
experimental data (Fig. \ref{fig4.5}A).

\begin{figure} [t]
\begin{center}
\includegraphics[width=12cm]{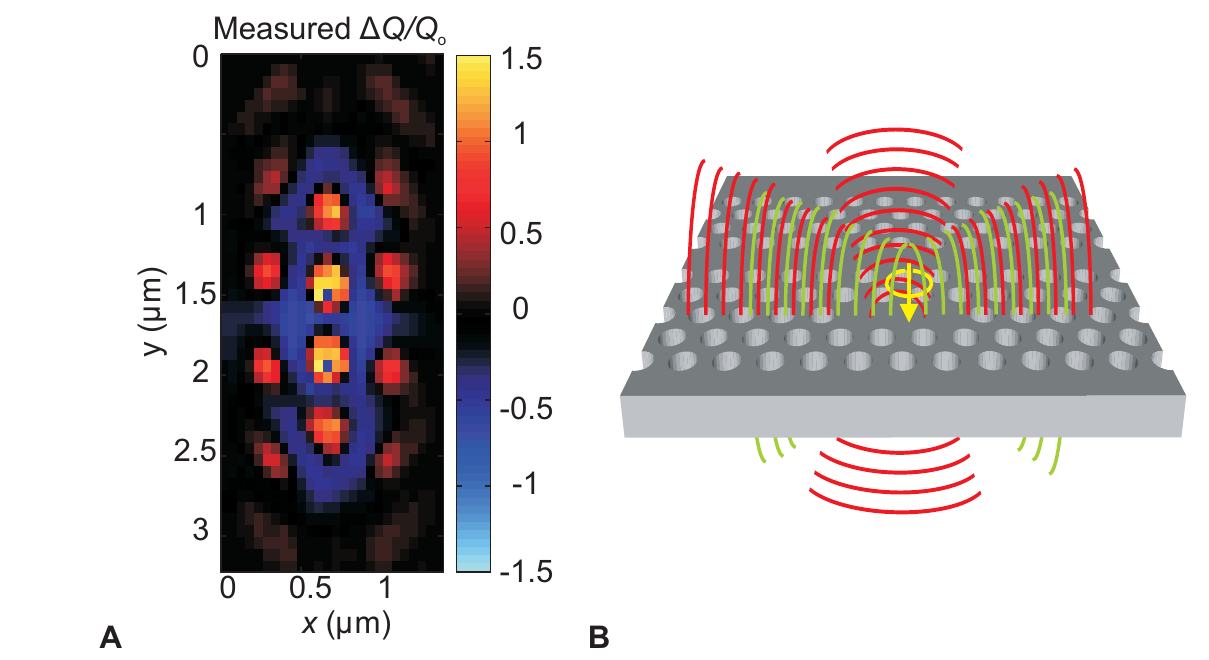}
\caption{{\rm \textbf{A}}, Image of the measured shift of the
quality factor $\mathrm{\Delta}Q/Q_\mathrm{o}$ at every probe
position. An increase and the decrease of the quality factor of the
cavity is evident. {\rm \textbf{B}}, Representation of the proposed
mechanism that causes the increase of the quality factor. As the
probe is above a maximum of the magnetic field, the induced current
generates radiation that destructively interferes with the radiative
loss of the cavity, yielding an increase of the photon lifetime.}
\label{fig4.6}
\end{center}
\end{figure}

In addition to tuning the resonance frequency, we also control the
lifetime of the photons in the cavity. In Fig. \ref{fig4.6}A, we
show an image of the retrieved $\mathrm{\Delta} Q/Q_\mathrm{o}$ as a
function of the probe position. Remarkably, the relative change in
$Q$ can, depending on the probe position, be both positive and
negative! The largest increase of $Q$ of 50\% occurs right above the
amplitude maxima of $B_z$, co-located with the largest blue-shift.
The magnetic coupling between the probe and the cavity, thus, not
only induces a novel blue-shift of the resonance but also causes the
photon lifetime in the cavity to be increased.

Any 2D cavity is affected by losses due to intrinsic out-of-plane
radiation \cite{akahane_high-q_2003}. In a previous study,
Robinson \emph{et al} \cite{robinson_far-field_2008} reported an
increase of only 1\% of $Q$ of a nanocavity which resulted from
the destructive interference between the out-of-plane radiation
and its back-reflection from a metallic object much larger than
the nanocavity itself. However, to significantly improve the $Q$,
one has to destructively interfere with a larger amount of the
out-of-plane radiation. We achieve 50\% increase of the photon
lifetime by exploiting the emission caused by the magnetic dipole
moment of the probe.  In fact, this induced dipole moment emits
primarily along the surface of the sample in counter phase with
respect to the driving field inside the cavity. On the other hand,
as shown by Fourier analyses performed on the cavity mode
\cite{akahane_high-q_2003}, the cavity also radiates along the
surface. We detected this radiation during a near-field
measurement, as indicated by the green arrows in Fig.
\ref{fig4.2}A where. Therefore, the in-plane radiation of the
cavity and the emission from the probe destructively interfere
(the process is schematically described in Fig. \ref{fig4.6}B).
Moreover, analyses on the in-plane decay rate show that we increase the quality of the cavity by also
decreasing the coupling with the access waveguide. Remarkably, we
obtain a pronounced increase of $Q$ by exploiting at the
nano-scale the scattered light from an object smaller than the
nanocavity. Furthermore, we achieve an increase of the lifetime by
means of magnetic coupling rather than electric.

Here, we have experimentally demonstrated that we can actively and
reversibly control the trapping of light in a photonic crystal
nanocavity by means of magnetic coupling with an actuated
subwavelength object. The presented method opens up a new way for
light control, combining photonic crystals and metamaterials
concepts.  Moreover, a new exciting application for photonic crystal
nanocavities arises. We anticipate the possibility of measuring the
magnetic dipole moment of magnetically resonant nano-object, such as
single split-ring resonator \cite{husnik_absolute_2008} or single
twisted split-ring resonator dimers
\cite{liu_stereometamaterials_2009}, by actuating it above a maximum
of the magnetic field of the nanoresonator.  Along these lines, we
also envision the striking possibility of using a state-of-the-art
ultra-high-$Q$ \cite{takahashi_high-q_2007} nanocavity,
characterized by a sharp resonance, for measuring the minute
magnetic susceptibility of molecules, such as carbon nanotubes
\cite{minot_determination_2004} or ring-shape (aromatic) molecules
\cite{haddon_magnetism_1995}.

\begin{acknowledgements}
This manuscript is extracted by the Ph.D. dissertation of M. Burresi. We wish to thank H. Schoenmaker for technical support and M. Bonn and G. H. Koenderink
for helpful discussion and support. This work is part of the
research program of the "Stichting voor Fundamenteel Onderzoek der
Materie (FOM)", which is financially supported by the "Nederlandse
organisatie voor Wetenschappelijk Onderzoek (NWO)". Support by the
NWO (VICI grant) is gratefully acknowledged. This work is also
supported by NanoNed, a nanotechnology program of the Dutch
Ministry of Economic affairs.
\end{acknowledgements}


%
%
\end{document}